\documentclass[AMA,STIX1COL, doublespace]{WileyNJD-v2}
\usepackage{moreverb}
\usepackage{amsmath}
\usepackage{xr-hyper}
\usepackage{hyperref}
\usepackage{dsfont}

\newcommand\BibTeX{{\rmfamily B\kern-.05em \textsc{i\kern-.025em b}\kern-.08em
T\kern-.1667em\lower.7ex\hbox{E}\kern-.125emX}}

\newcommand{\probP}{\text{I\kern-0.15em P}}

\articletype{}%

\begin{document}

\title{Augmented Binary Method for Basket Trials (ABBA)}

\author[1]{Svetlana Cherlin*}

\author[1]{James M.S. Wason}

\authormark{Svetlana Cherlin \textsc{et al}}

\address[1]{\orgdiv{Population Health Sciences Institute}, \orgname{Newcastle University}, \orgaddress{\state{ Ridley Building 1, Newcastle upon Tyne}, \country{UK}}}

\corres{*Svetlana Cherlin, Population Health Sciences Institute, Newcastle University, Ridley Building, Newcastle upon Tyne, UK. \email{svetlana.cherlin@newcastle.ac.uk}}

\abstract[Abstract]{In several clinical areas, traditional clinical trials often use a responder outcome, a composite endpoint that involves dichotomising a continuous measure. An augmented binary method that improves power whilst retaining the original responder endpoint has previously been proposed. The method leverages information from the the undichotomised component to improve power. We extend this method for basket trials, which are gaining popularity in many clinical areas. For clinical areas where response outcomes are used, we propose the new Augmented Binary method for BAsket trials (ABBA) enhances efficiency by borrowing information on the treatment effect between subtrials. The method is developed within a latent variable framework using a Bayesian hierarchical modelling approach. We investigate the properties of the proposed methodology by analysing point estimates and credible intervals in various simulation scenarios,  comparing them to the standard analysis for basket trials that assumes binary outcome. Our method results in a reduction of 95\% high density interval of the posterior distribution of the log odds ratio and an increase in power when the treatment effect is consistent across subtrials. We illustrate our approach using real data from two clinical trials in rheumatology.}


%\jnlcitation{\cname{
%\author{Cherlin S.}, 
%\author{Wason J. M. S.}} 
%\ctitle{Augmented Binary Method for Basket Trials (ABBA)}, \cvol{2024}.}

\maketitle

\section{Introduction}\label{intro}

In many clinical trials for complex diseases, selecting a single primary endpoint has proven challenging \cite{wason:etal:2020}. Responder endpoints are commonly proposed to combine distinct components into a single measure, with at least one of them often being continuous. These continuous components are often dichotomised,  meaning that classifying a patient as a responder depends on achieving a pre-specified threshold, in addition to meeting specific criteria for the binary component(s). For example, in the context of cancer trials, a patient is defined as a responder if their tumour size shrinks by more than 30\% and they do not have new tumour lesions detected \cite{eisenhauer:etal:2009}. However, it has been shown that using binary responder endpoints leads to larger expected and maximum sample sizes compared to continuous endpoints \cite{wason:etal:2011}. To this end, the augmented binary method \cite{wason:seaman:2013} has been developed to model the joint distribution of the continuous outcome and a binary outcome. The method improves the precision of analysis and increases power by avoiding the loss in information from dichotomising a continuous component. Initially developed for cancer trials, the method has since been adapted for use in clinical trials for Rheumatoid Arthritis  \cite{wason:jenkins:2016}. 

In the area of rheumatic diseases, the most common outcomes are the Disease Activity Score, which includes different 28-joint counts (DAS28)\cite{felson:etal:1995}, and the American College of Rheumatology (ACR) improvement criterion \cite{siegel:zhen:2005}; both are based on dichotomised information such as DAS28 being less than a specific threshold, or ACR achieving a pre-specified percentage threshold. Additionally, a binary indicator $b$ of whether or not the patients received rescue medications or discontinued treatment is often used, known as `non-responder imputation'. Patients are considered responders based on both binary or dichotomised continuous variables, and the binary indicator $b$. The augmented binary method does not dichotomise the outcome measure. Instead, it analyses the continuous outcome (DAS28 or ACR), and the binary indicator $b$ described above. By jointly modelling continuous and binary outcomes, the method utilises information on the distance from the dichotomisation threshold for the continuous outcome. The augmented binary method outperformed the standard method that analyses binary endpoints in Rheumatoid Arthritis, as evident by a substantial reduction in the 95\% confidence intervals for the probability of success without inflation in type I error rate, indicating that the estimated probability of response is more precisely estimated. The augmented binary method has been extended to incorporate multiple follow up times \cite{lin:wason:2016} and multiple doses of the same treatment within the latent variable framework \cite{wason:seaman:2020, mcmenamin:etal:2020}. Additionally, the original augmented binary method has been implemented in a Bayesian framework \cite{trialr}.

However, the method has so far been restricted to standard single-arm and two-arm clinical trials. Recently, basket trials, which aim to investigate a single therapy across multiple conditions, have become popular\cite{hobbs:etal:2022}. Basket trials consist of subtrials characterised by a specific disease, with each subtrial investigating the same treatment. Originating in the field of oncology and aiming to target shared molecular aberrations, basket trials have been extended to other areas such as immune-mediated inflammatory diseases, where similar symptoms are shared across multiple conditions\cite{grayling:etal:2021}. The efficiency of basket trials lies in their requirement of fewer patients and a shorter duration compared to traditional trial designs, owing to the borrowing of information between subtrials. It has been demonstrated that sharing information between the subtrials in a basket trial leads to  more efficient utilisation of data \cite{berry:etal:2013, zheng:wason:2022, ouma:etal:2022, whitehead:etal:2023, zhou:ji:2024}.

In this work, we extend the augmented binary method to basket trials. The method is based on a latent variable approach and is implemented in a Bayesian framework. It utilises a multivariable normal distribution to model the continuous outcome and a latent variable that is assumed to determine the binary indicator $b$. By avoiding dichotomisation of the outcome, we aim to achieve more precise inference and increased power of the trial. We illustrate the method using clinical trials in rheumatology where the same responder endpoint is used. Additionally, the method could be applicable beyond rheumatology, such as in oncology trials where response rate is used as an endpoint. 

\nocite{*}
\section{Methods}\label{methods}
\subsection{Notation}

Suppose a randomised, placebo-controlled trial is conducted, in which patients $i = 1, \dots, N$ are equally randomised to treatment ($t_{i} = 1$) or control ($t_{i} = 0$). 
At the end of the trial, two outcomes are measured for each patient: a continuous score, such as a DAS28 or an ACR-N score  ($y_{i1}$) and a binary indicator for receiving rescue medication ($y_{i2}$). 

We assume that we are working with the ACR20 response criteria, where a response is defined if the ACR-N is dichotomised at 20\%. We work on the log scale, as this was shown to be closer to normally distributed in previous work \cite{wason:jenkins:2016}. According to this criteria, a patient is defined as a responder if
\begin{equation*}
y_{i1} \ge \mathrm{log}(20)~\mathrm{and}~y_{i2} = 0. 
\end{equation*}
We assume that there is a continuous latent variable $y_{i2}^\star$  that determines whether a patient receives rescue medication or is a non-responder for another pre-defined reason: 
\begin{equation*}
y_{i2} = \mathds{1}(y_{i2}^\star > 0),
\end{equation*}
where $\mathds{1}$ is an indicator function taking the value 1 if its argument is true and 0
otherwise. Thus, the patient is a responder if 
\begin{equation*}
y_{i1} > \mathrm{log}(20)~\mathrm{and}~y_{i2}^\star > 0.
\end{equation*}

\subsection{Stratified Augmented Binary Method}

We introduce an augmented binary method for a non-basket framework based on a latent variable approach (hereafter referred to as \textit{stratified analysis}). The outcomes are modelled using a multivariate probit regression model as follows:
\begin{align}
y_{i1} &= \mu_{i1} +\epsilon_{i1}, ~y_{i2}^\star = \mu_{i2} +\epsilon_{i2} \nonumber \\
(\epsilon_{i1}, \epsilon_{i2})^T &\sim MVN\{0, \boldsymbol{\Sigma});~
\boldsymbol{\Sigma}= \left( \begin{array}{ccc}
\sigma_1^2 & \sigma_1\sigma_2\rho \\  
\sigma_2\sigma_1\rho & \sigma_2^2\end{array} \right).  \label{eq:outcomes} 
\end{align}
We decompose the covariance matrix $\boldsymbol{\Sigma}$ as $\boldsymbol{\Sigma} = \mathrm{diag}(\boldsymbol{\sigma}) \times \boldsymbol{\Omega} \times \mathrm{diag}(\boldsymbol{\sigma})$,
where $\boldsymbol{\Omega}$ is the correlation matrix
$\boldsymbol{\Omega}= \left( \begin{array}{ccc}
\rho & 1\\
1 & \rho \end{array} \right)$ and $\boldsymbol{\sigma}$ is the vector of standard deviations $\boldsymbol{\sigma} = (\sigma_1, \sigma_2)$. For details on implementing the multivariate probit regression model approach, see Section \ref{sec:implementation}. The standard deviation of the latent variable, $\sigma_2$, is set to 1 for identifiability reasons.

The means of the outcomes are modelled as follows:
\begin{align*}
\mu_{i1} &= \beta_{1} + \gamma_{1}\mathrm{log}(x_{i}) + \theta_{1}t_{i}, \\
\mu_{i2} &= \beta_{2} + \gamma_{2}\mathrm{log}(x_{i}) + \theta_{2}t_{i},
\end{align*} where $x_i$ is the baseline measure for some disease activity score or any other clinically important covariate for patient $i$, 
$\boldsymbol\beta = (\beta_{1}, \beta_{2})$, $\boldsymbol\gamma = (\gamma_{1}, \gamma_{2})$ and $\boldsymbol\theta = (\theta_{1}, \theta_{2})$ are vectors of the intercepts, the effects of the baseline measure and the treatment effect, respectively. 

Based on previous research \cite{zheng:wason:2022}, the parameters for the intercept and baseline effect, $\boldsymbol\beta = (\beta_{1}, \beta_{2})$, $\boldsymbol\gamma = (\gamma_{1}, \gamma_{2})$ are assigned independent normal priors $N(0, 5^2)$. The treatment effects, $\boldsymbol\theta = (\theta_{1}, \theta_{2})$, are assigned independent normal priors $N(0, 10^2)$. The standard deviation of the continuous outcome is given an inverse gamma prior $\sigma_1 \sim IG(0.5, 0.005)$. The correlation matrix is given an LKJ prior\cite{lewandowski:etal:2009} $\boldsymbol\Omega \sim \mathrm{LKJCorr}(\eta)$ with a shape parameter $\eta$ = 5. The shape parameter was chosen to allow for a wide range of correlations.

\subsection{Augmented Binary Method for Basket Trials (ABBA)}

We now extend the method to cover a randomised, placebo-controlled basket trial with $K$ subtrials. 
We assume a total of $N$ patients are included in the study with $K$ subtrials ($i = 1, \dots, N)$. Within each subtrial $k = 1, \dots, K$, the patients are equally randomised to treatment ($t_{i} = 1$) or control ($t_{i} = 0$), and the outcomes are collected and modelled similarly to the stratified augmented binary model (see Equation \eqref{eq:outcomes}).

The means of the outcomes are modelled as follows:
\begin{align*}
\mu_{i1} &= \beta_{k1} + \gamma_{k1}\mathrm{log}(x_{i}) + \theta_{k1}t_{i}, \\
\mu_{i2} &= \beta_{k2} + \gamma_{k2}\mathrm{log}(x_{i}) + \theta_{k2}t_{i},
\end{align*} where 
$\boldsymbol\beta_{k} = (\beta_{k1}, \beta_{k2})$, $\boldsymbol\gamma_{k} = (\gamma_{k1}, \gamma_{k2})$ and $\boldsymbol\theta_{k} = (\theta_{k1}, \theta_{k2})$ are vectors representing the intercepts, the effects of the baseline measure, and the treatment effect for subtrial ${k}$. 

\subsubsection{Prior Level 1}

The vectors  $\boldsymbol\beta_{k} = (\beta_{k1}, \beta_{k2})$, $\boldsymbol\gamma_{k} = (\gamma_{k1}, \gamma_{k2})$ and $\boldsymbol\theta_{k} = (\theta_{k1}, \theta_{k2})$ for subtrial $k$ have independent multivariate normal priors:

\begin{align*}
\boldsymbol\beta_{k} &\sim MVN(\boldsymbol\mu_\beta, \boldsymbol\Sigma_\beta),\\
\boldsymbol\gamma_{k} &\sim MVN(\boldsymbol\mu_\gamma, \boldsymbol\Sigma_\gamma),\\
\boldsymbol\theta_{k} &\sim MVN(\boldsymbol\mu_\theta, \boldsymbol\Sigma_\theta),\\
\end{align*} where 
\begin{align*}
\boldsymbol{\Sigma_\beta}= \left( \begin{array}{ccc}
\sigma_{\beta1}^2 & 0  \\
0 & \sigma_{\beta2}^2\end{array} \right),
\boldsymbol{\Sigma_\gamma}= \left( \begin{array}{ccc}
\sigma_{\gamma1}^2 & 0  \\
0 & \sigma_{\gamma2}^2\end{array} \right),
\boldsymbol{\Sigma_\theta}= \left( \begin{array}{ccc}
\sigma_{\theta1}^2 & 0  \\
0 & \sigma_{\theta2}^2\end{array} \right).
\end{align*}

\subsubsection{Prior Level 2}

The standard deviations $\boldsymbol\sigma_\beta = (\sigma_{\beta1}, \sigma_{\beta2})$,  $\boldsymbol\sigma_\gamma = (\sigma_{\gamma1}, \sigma_{\gamma2})$ and $\boldsymbol\sigma_\theta = (\sigma_{\theta1}, \sigma_{\theta2})$ have independent exponential priors $Exp(2)$. Assuming similar effects across the subtrials, the prior was chosen to yield a 0.86 probability that each parameter is less than 1, which is suitable when modelling the continuous outcome on the log scale. These parameters determine the extent of information sharing between subtrials. The lower boundaries for $\boldsymbol\sigma_\beta$, $\boldsymbol\sigma_\gamma$ and $\boldsymbol\sigma_\theta$ are constrained to 0.1 (\i.e.$Exp(2)$ truncated at 0.1), to prevent convergence issues. Similarly to the stratified analysis, the prior means $\boldsymbol\mu_\beta = (\mu_{\beta1}, \mu_{\beta2})$, $\boldsymbol\mu_\gamma = (\mu_{\gamma1}, \mu_{\gamma2})$ and $\boldsymbol\mu_\theta = (\mu_{\theta1}, \mu_{\theta2})$ have independent normal prior distributions $N(0,5)$. The standard deviation of the continuous outcome is assigned an inverse gamma prior $\sigma_1 \sim IG(0.5, 0.005)$. The correlation matrix is assigned an LKJ prior $\boldsymbol\Omega \sim \mathrm{LKJCorr}(\eta)$ with a shape parameter $\eta$ = 5.

\subsection{Binary Method for Basket Trials (BIN)}

We compare the ABBA method with the BIN method, where we fit a logistic regression model to the dichotomised responder data, modelling the probability of response with the logistic regression
\begin{align}
\mathrm{logit}(p_{i}) =  \xi_i,  \label{eq:bin}
\end{align} 
where $\xi_i =  \beta + \gamma\mathrm{log}(x_{i}) + \theta t_{i}$. The parameters $\beta$, $\gamma$ and $\theta$ have independent normal priors $N(0,5)$. For a basket trial, the BIN method models the binary responder data with a logistic regression that allows for sharing of information between  subtrials:
\begin{align*}
\xi_i = \beta_{k} + \gamma_{k}\mathrm{log}(x_{i}) + \theta_{k}t_{i},  
\end{align*}
where $\beta_{k}$, $\gamma_{k}$ and $\theta_{k}$ are the intercept, baseline effect and treatment effect for subtrial $k$.
The priors for these parameters are:
\begin{align*}
\boldsymbol\beta_{k} &\sim N(\mu_\beta, \sigma_\beta),\\
\boldsymbol\gamma_{k} &\sim N(\mu_\gamma, \sigma_\gamma),\\
\boldsymbol\theta_{k} &\sim N(\mu_\theta, \sigma_\theta).\\
\end{align*}
The prior means $\mu_\beta$, $\mu_\gamma$ and $\mu_\theta$ have independent normal priors $N(0,5)$.
The standard deviations $\sigma_{\beta}$, $\sigma_{\gamma}$ and $\sigma_\theta$, which control the amount of information sharing between subtrials, have independent exponential priors $Exp(2)$ truncated at 0.3, due to convergence issues.

\subsection{Inferring success probability}
For the ABBA method, the probability of success for participant $i$ is the probability of being a responder inferred from the model:
\begin{align*}
\probP(y_{i1} > \mathrm{log}(20)~\mathrm{and}~y_{i2}^\star > 0|\boldsymbol\nu) = 
\int_{\mathrm{log}(20)}^{\infty} \int_{0}^{\infty}f_{Y_1, Y^\star_2}(y_{i1}, y_{i2}^\star; \boldsymbol\nu)\mathrm{d}y_{i1}\mathrm{d}y_{i2}^\star,
\end{align*}

where $f_{Y_1, Y^\star_2}(y_{1i}, y_{2i^\star}; \boldsymbol\nu)$ is  the pdf of the bivariate normal distribution from Equation \eqref{eq:outcomes}, and $\boldsymbol\nu$ is a vector of model parameters $\nu = (\boldsymbol{\beta}, \boldsymbol{\gamma}, \boldsymbol{\theta}, \boldsymbol{\mu_\beta}, \boldsymbol{\mu_\gamma}, \boldsymbol{\mu_\theta}, \boldsymbol{\sigma_\beta}, \boldsymbol{\sigma_\gamma}, \boldsymbol{\sigma_\theta},\boldsymbol{\Omega}, \sigma_1)$.

The log odds ratio for the $k$-th subtrial, $\lambda_k$, is computed  as
\begin{align*}
\lambda_k = \mathrm{log}\left( \frac{(r_{k,t})/(1-r_{k,t})}{(r_{k,c})/(1-r_{k,c})} \right),   
\end{align*}
where $r_{k,t}$ and $r_{k,c}$ are the means of the success probabilities in the treatment and control groups for subtrial $k$, respectively.

For the BIN method, the probability of success for participant $i$ is computed as follows:
\begin{align*}
\probP(y_{i} = 1|\boldsymbol\nu)  = exp(\xi_i|\boldsymbol\nu)/(1 + exp(\xi_i|\boldsymbol\nu)) 
\end{align*}
where $\xi_i$ is the linear predictor from Equation \eqref{eq:bin}, and $\boldsymbol\nu$ is a vector of model parameters $\boldsymbol\nu = (\boldsymbol{\beta}, \boldsymbol{\gamma}, \boldsymbol{\theta}, \mu_\beta, \mu_\gamma, \mu_\theta, \sigma_\beta, \sigma_\gamma, \sigma_\theta)$.

\subsection{Implementation}\label{sec:implementation}
The Bayesian analysis models are implemented in Stan \citep{stan}, using the ``rstan'' package \cite{rstan}. For the multivariate probit regression, the latent continuous variable is divided into two parts based on whether the corresponding observed binary variable is 0 or 1. This division results in positive-constrained and negative-constrained components, where the sizes correspond to the counts of true (1) and false (0) observations in the binary variable.
The positive-constrained and negative-constrained components are sampled separately from the uniform distribution, and the latent variable is constructed to account for the indices of 0s and 1s in the observed binary variable. To obtain the posterior distribution of the parameters, we employ two parallel chains, each running the Hamiltonian Monte Carlo sampler for 10,000 iterations after a burn-in period of 5,000 iterations. \texttt{R} code for implementing the ABBA and BIN methods (with and without borrowing of information) is available on  GitHub (\url{https://github.com/svetlanache/ABBA}).

\section{Simulation study}

To understand the properties of the ABBA model and compare it with the BIN model, we conducted a series of simulation studies. In all scenarios, we simulated basket trials with 3 subtrials, each having a sample size of 50 participants, with equal randomisation (25 participants in each arm). We chose this relatively small sample size to reflect the typical sample sizes observed in real-world phase II clinical trials. To investigate the method's performance with larger samples, we also considered a sample size of 100 participants per arm for one of the scenarios. 
Working with the posterior distribution of the log odds ratios $\boldsymbol{\lambda}$ = $(\lambda_1, \dots, \lambda_K)$, where $K$ is the number of subtrials, we evaluate the following performance measures as recommended in a recent tutorial  \cite{morris:etal:2019}: 

(i) bias of the posterior means of  $\boldsymbol{\lambda}$, defined as the average distance from the true value of the log odds ratios, averaged across simulation replicates:
\begin{align*}
\mathrm{Bias}(\boldsymbol{\lambda})  = \frac{1}{M}\sum_{m=1}^{M} \boldsymbol{\bar{\lambda}_m} - \boldsymbol{\lambda_m},
\end{align*}
where $M$ is the total number of simulation replicates, and $\boldsymbol{\bar{\lambda}_m}$ is the posterior mean of the log odds ratio for the $m$-th replicate.

(ii) posterior precision of $\boldsymbol{\lambda}$, defined as the reciprocal of the posterior variance of $\boldsymbol{\lambda}$, averaged across simulation replicates:
\begin{align*}
\mathrm{Precision}(\boldsymbol{\lambda})  = \frac{1}{M}\sum_{m=1}^{M} \frac{1}{\mathrm{Var}(\boldsymbol\lambda_m)}.
\end{align*}
To evaluate the properties of the 95\% high density interval (HDI) for $\boldsymbol{\lambda}$, we consider:

(iii) one-sided power defined as a the proportion of simulations where the lower limit of 95\% HDI for $\boldsymbol{\lambda}$ is greater than 0, averaged across simulation replicates:
\begin{align*}
\frac{1}{M}\sum_{m=1}^{M} \mathds{1} (\boldsymbol{\lambda_{m, low}} > 0)
\end{align*}
(iv) width of 95\% HDI for $\boldsymbol{\lambda}$, averaged across simulation replicates:
\begin{align*}
\frac{1}{M}\sum_{m=1}^{M} (\boldsymbol{\lambda_{m, up}} - \boldsymbol{\lambda_{m, low}})
\end{align*}
(v) coverage, defined as the proportion of simulation replicates in which the 95\% HDI for  $\boldsymbol{\lambda}$ includes the true value:

\begin{align*}
\frac{1}{M}\sum_{m=1}^{M} \mathds{1} (\boldsymbol{\lambda_{m,low}} \leq \boldsymbol{\lambda_k} \leq \boldsymbol{\lambda_{m,up}}).
\end{align*}
We investigated the following scenarios:\\
Scenario 1: Null scenario (no treatment effect).\\
Scenario 2: Treatment effect on the continuous component only, consistent across subtrials.\\
Scenario 3: Treatment effect on the latent component only, consistent across subtrials.\\
Scenario 4: Treatment effect on both components, consistent across subtrials.\\
Scenario 5: Treatment effect on both components consistent in two subtrials, no treatment effect in the other subtrial.\\
Scenario 6: Treatment effect on both components in one subtrial, no treatment effect in the other two subtrials.\\
Scenario 7: Treatment effect consistent on the continuous component, inconsistent on the latent component.\\
Scenario 8: Treatment effect consistent on the latent component, inconsistent on the continuous component.\\
For each scenario, we simulated 1,000 replicates. (For the null scenario, we simulated 5,000 replicates. For a true type I error rate of 0.05, this would produce a Monte Carlo standard error for the estimated type I error rate of 0.003.) Table \ref{tab:simulations_scenarios} presents the parameters used in the simulations, and the characteristics of the simulated data for the eight scenarios mentioned above.
To simulate the data, the parameter vectors for the baseline effect $\boldsymbol{\gamma}$ and the treatment effect $\boldsymbol{\theta}$ were fixed. The vectors of intercepts $\boldsymbol{\beta}$ were optimised to achieve a target response rate, which was 0.2 for the null scenarios, and 0.2 and 0.4 for the control and treatment arms, respectively, in the non-null scenarios (optimisation was performed using the \texttt{optim} function in \texttt{R}). We used a standard deviation of $\sigma_1$ = 0.5 for the continuous outcome, $\sigma_2$ = 1 for the latent variable, and a correlation parameter $\rho$ = 0.3. To compare the proposed methodology with the analysis that does not borrow information between  subtrials, we also conducted a stratified analysis, examining each subtrial separately for scenarios presented in Table {\ref{tab:simulations_scenarios}}.

\section{Results} 

Table \ref{tab:results_basket} summarises the width of the 95\% HDI for the log odds ratio and the power for scenarios presented in Table \ref{tab:simulations_scenarios}.
The ABBA method demonstrates a reduction of 20\% - 26\% in the width of the 95\% HDI for the log odds ratio, and an increase in power of 25\% - 70\%.
For Scenario 4, we simulated additional data with 100 participants per arm and achieved powers of 0.79, 0.78, 0.77 for the BIN model. Similar powers (0.75,0.75 and 0.74) were achieved with the ABBA model using 50 participants per arm. Thus, the increase in power with the ABBA model is equivalent to a 50\% reduction in sample size. Although the type I error rate appears to increase for the ABBA method, this result is misleading as the type I error rate remains well controlled at 5\%.

To evaluate the properties of the ABBA and BIN models without borrowing information between subtrials, we conducted separate analyses for each subtrial (stratified analysis) for scenarios presented in Table {\ref{tab:simulations_scenarios}}. Table \ref{tab:results_stratified} presents the 95\% HDI for the log odds ratio and the power. The ABBA method results in a reduction of  25\% - 32\% in the width of the 95\% HDI across different scenarios, and an increase in power of 30\% - 75\%. The type I error is controlled at 5\%.

Table \ref{tab:ab_abba} compares the ABBA method with and without sharing of information. Sharing information between subtrials results in a reduction of 15\% - 20\% in the width of the 95\% HDI for the log odds ratio. There is a notable decrease in type I error rate for Scenario 1 and an increase in power for most scenarios. Interestingly, subtrial 3 in Scenario 6 and subtrial 3 in Scenario 7 show a reduction in power for the ABBA method that shares information between subtrials, despite the reduction in the 95\% HDI. In these scenarios, there is a treatment effect in subtrial 3, whereas there is no treatment effect (Scenario 6), or inconsistent treatment effect (Scenario 7) in the other two subtrials. In scenario 6, sharing information results in a slight inflation in type I error rate (6\%). Additionally, the smallest increase in power is observed for Scenario 5 (subtrials 1 and 2), and Scenario 8 (subtrial 3), where the treatment effect is not consistent for all  subtrials.

\begin{figure*}[t]

	\includegraphics[scale=1]{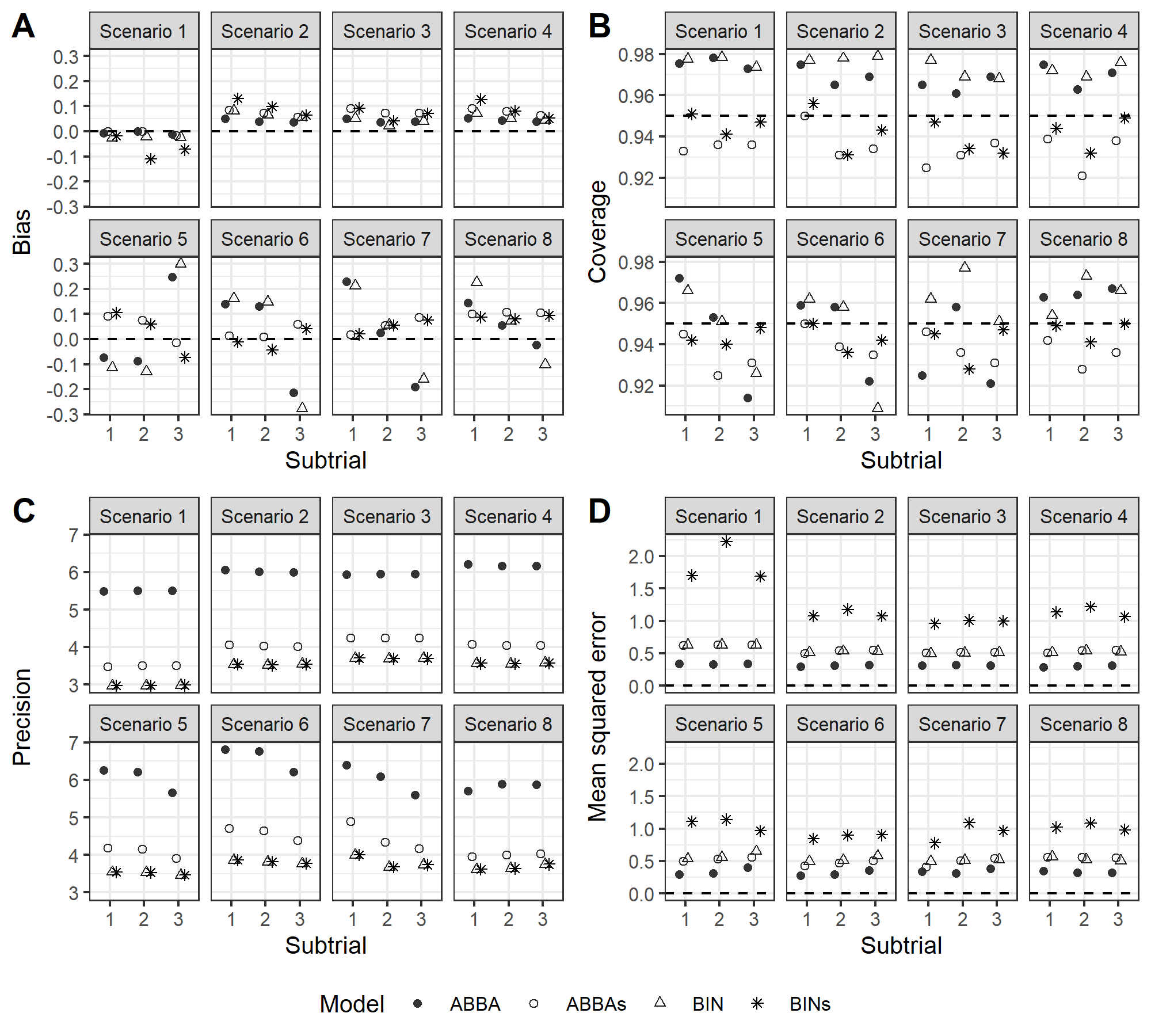}
	\caption{Bias, mean squared error and coverage of the posterior estimators for the log odds ratios.\label{fig:sim_results}}
\end{figure*}

Figure \ref{fig:sim_results} compares the bias (panel A), precision (panel B) and coverage (panel C) of the posterior mean for the log odds ratio, for the ABBA and BIN models with and without information sharing (the methods without information sharing are denoted as ABBAs and BINs to indicate stratified analysis). It demonstrates that the ABBA model with information sharing produces equivalent or smaller bias across nearly all scenarios (panel A), and substantially higher precision in all scenarios (panel B). The coverage for the ABBA method (panel C) is above nominal for Scenario 1-4 where the treatment effect is consistent across subtrials. However, for Scenario 5-7, the coverage is below nominal for some subtrials. The lowest coverage is 91.4\% for scenario 5, subtrial 3, where there is no treatment effect, while the treatment effect is consistent across both components in the other two subtrials. In Scenario 6, the coverage is below nominal for subtrial 3, which differs from the other subtrials in terms of both components of the treatment effect. For Scenario 7, the coverage is below nominal for subtrials 1 and 3 where the treatment effects for the latent component have opposite directions, despite a consistent effect on the continuous component. Consistent treatment effect on the latent component only (Scenario 8) does not affect the coverage of the ABBA method, whereas lack of information sharing (ABBAs method) results in coverage slightly below nominal.

\section {Real Data Analysis}

To illustrate the performance of the models on real data, we conducted a re-analysis of data from two clinical trials. This study, carried out under YODA Project 2023-5145, used data obtained from the Yale University Open Data Access Project, which has an agreement with Janssen Research \& Development, L.L.C.. The interpretation and reporting of research using this data are solely the responsibility of the authors and does not necessarily represent the official views of the Yale University Open Data Access Project or Janssen Research \& Development, L.L.C.. In both clinical trials, we examined outcomes related to DAS28 and ACR-N. 

In  NCT01645280 (EudraCT NUMBER: 2011-001122-18, Protocol CNTO1275ARA2001) \cite{smolen:etal:2017}, patients were randomised into five groups: placebo (n=55), two groups receiving ustekinumab (different regiments, n=55 each), and guselkumab (different regiments, n = 55 and 54). The primary endpoint was the proportion of patients achieving at least a 20\% improvement in the American College of Rheumatology criteria (ACR20) at week 28, \i.e. the proportion of responders according to the ACR20 criteria. Patients were classified as non-responders if no ACR component data were available at week 28 or if they initiated prohibited medications (including glucocorticoids for Rheumatoid Arthritis), increased the methotrexate or glucocorticoid dose above the baseline level, or discontinued the study agent for any reason.

We compared the Disease Activity Score using CRP (DAS28-CRP)  at 28 weeks between usetkinumab (combining two dose groups) and placebo arms in the primary analysis population. If patients had data for at least one DAS28-CRP component at week 28, missing components were imputed using the last observation carried forward method. In the latent variable model, DAS28-CRP at week 28 represented a continuous component of the outcome, while an indicator of drug withdrawal or administration of prohibited medications represented a binary component. In the binary model, patients were classified as responders if the drug was not withdrawn during the study, no prohibited medications were administered, and DAS28-CRP at week 28 was below the threshold of 2.6. We adjusted for DAS28-CRP at baseline in the analyses.
For the ACR-N outcome, we computed the ACR-N score at 28 weeks and adjusted the analysis for the baseline value of the DAS28-CRP measure. In  the binary model for ACR-N, patients were classified as responders if there ACR-N score exceeded 20\%. 

In NCT01077362 (EudraCT NUMBER: 2009-012265-60 , Protocol CNTO1275PSA3002, \cite{ritchilin:etal:2014}), 312 adults with active psoriatic arthritis were randomised to different doses of ustekinumab or placebo with  crossover to ustekinumab. We compared DAS28-CRP at week 24 between ustekinumab and placebo arms in the primary analysis population.
The outcomes were specified similarly to those for the NCT01645280 trial (see Table \ref{tab:realdata}).

The resulting 95\% HDI for the log odds ratios are presented in Table \ref{tab:resultsreal} for a hypothetical basket trial that combines the two trials (combined analysis), and for each trial analysed separately (stratified analysis).  For NCT01645280 and NCT01077362, sharing of information in the ABBA model resulted in 5\% and 6\% decrease in the width on 95\% HDI for the DAS28 outcome, and 4\% and 2\% for the ACR-N outcome, respectively. The ABBA model offers a substantial advantage in reducing the width of the 95\% HDI (16\% - 46\%), compared to the BIN model.\\

\section{Discussion}

In this paper, we have proposed and assessed the augmented binary method for basket trials, which  assumes an underlying continuous latent variable that determines the observed binary component, jointly modelling it with the observed continuous component. This method draws motivation from the augmented binary method which previously demonstrated superior performance in non-basket rheumatology trials compared to standard analyses of binary outcomes. We conducted various comparisons of our ABBA method against standard methods and demonstrated its substantial advantages: it achieves a notable reduction in the width of the 95\% HDI for the log odds ratio, along with increased power and precision compared to standard logistic regression methods that do not utilise the continuous nature of outcomes.

When comparing the ABBA method with information borrowing between subtrials to the method without borrowing, we observed a reduction of 17\% to 20\% in the width of the 95\% HDI across various simulation scenarios. Our simulation study further highlighted that the ABBA method exhibited the highest precision for the posterior mean of the log odds ratio compared to other methods. However, in scenarios where the treatment effect varied inconsistently across subtrials, the ABBA methods showed a slight reduction in power and a drop in coverage. This suggests that  careful consideration is necessary regarding the expected consistency of effects across subtrials. Nevertheless, even when the treatment effect was consistent for the latent component only and inconsistent for the continuous component, sharing of information through ABBA was advantageous. 

We illustrated the application of our method to real data from rheumatology trials. Although the trials were conducted separately we hypothesised a scenario resembling a basket trial comprising two subtrials due to the investigation of the same treatment and common clinical outcomes. We investigated how combining the two trials into a basket trial might have affected the results. The ABBA model achieved a substantial reduction in the width of the 95\% HDI compared to the analysis of binary outcomes with sharing information between subtrials. However, the reduction achieved by ABBA with information sharing was minor compared to that without sharing. This can be attributed to our hypothesised basket trial comprising only two subtrials, thereby limiting the amount of expected information sharing.

\begin{figure*}[t]
	\includegraphics[scale=0.6]{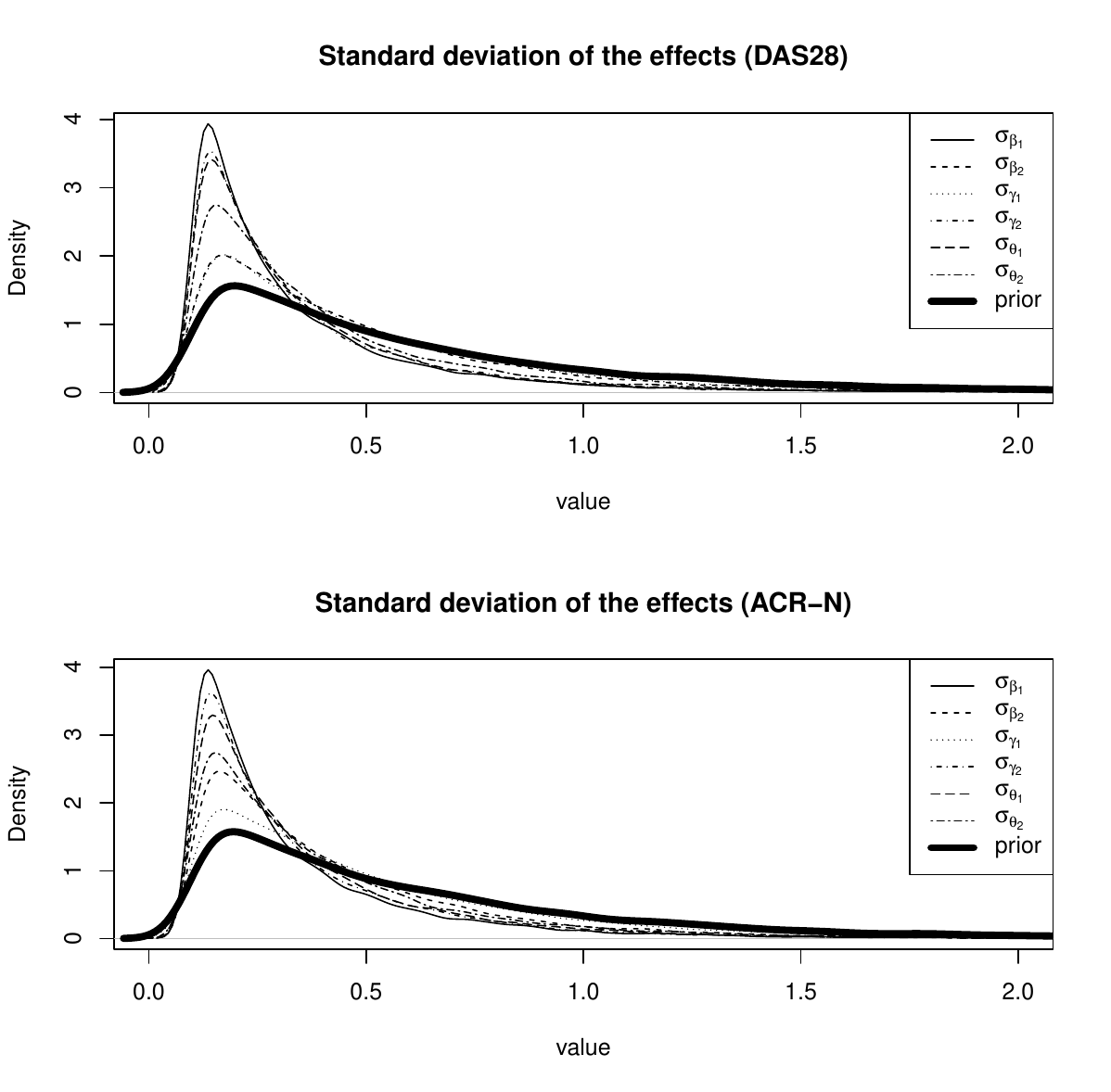}
	\caption{Posterior and prior distributions of the standard deviation of the effects.\label{fig:sd_results}}
\end{figure*}

Our method has a few limitations, such as longer computational time and occasional convergence issues of the MCMC algorithm. Another issue is a prior specification, especially for the variation between subtrials. With a small number of subtrials, the data contributes minimally to the posterior distribution which is primarily influenced by the prior (Figure \ref{fig:sd_results}). Thus, the choice of the prior for this parameter can significantly impact the degree of information borrowing across subtrials \cite{berry:etal:2013,cunanan:etal:2019,gelman:2006}. 
In this study, we utilised an exponential prior that assigns higher probability to smaller values of the parameter. Future research will explore alternative priors such as commensurate prior which relies on distributional discrepancy to measure commensurability between  subtrials and borrows information from those with the most similar treatment effect. \cite{zheng:wason:2022}. Our method assumes that the clinical outcome is the same in each disease. This is not always the case, as different conditions in a basket trial often have different clinical outcomes. However, in this case, there might be a common secondary outcome (e.g., a mechanistic biomarker) that is the same for each condition. This setting is particularly relevant, but not limited to, clinical trials in rheumatology. We therefore aim to investigate basket trials involving distinct outcomes where information may be shared via a common secondary outcome. 

In conclusion, the augmented binary method for basket trials makes more efficient use of data by borrowing information across subtrials, presenting a promising avenue for further research.

\bibliography{references}
\clearpage

% TABLES
%%%%%%%%%

%Table 1
\begin{table}[h!]
	\begin{tabular}{cc|c|c|c|c|c|c|c|c|c}
		Scenario & Subtrial &
		\multicolumn{1}{|p{2cm}|}{\centering  Mean LOR \\ (SD)} &
		\multicolumn{1}{|p{2cm}|}{\centering  Mean RR$_c$ \\ (SD)} &
		\multicolumn{1}{|p{2cm}|}{\centering  Mean RR$_t$ \\ (SD)} &
		$\beta_1$ & $\beta_2$ & $\gamma_1$ & $\gamma_2$ &$\theta_1$ & $\theta_2$  \\ \hline
		\multirow{3}{*}{1}  
		& 1 & 0(0.16) & 0.2(0.018) & 0.2(0.018) & 1.07 & 0.04 & 0.5 & -0.1 & 0 & 0 \\ 
		& 2 & 0(0.16) & 0.2(0.018) & 0.2(0.018) & 1.07 & 0.04 & 0.5 & -0.1 & 0 & 0 \\
		& 3 & 0(0.16) & 0.2(0.018) & 0.2(0.018) & 1.07 & 0.04 & 0.5 & -0.1 & 0 & 0 \\ 
		\hline
		\multirow{3}{*}{2}  
		& 1 & 0.99(0.17) & 0.19(0.02) & 0.39(0.02) & 0.84 & 0.39 & 0.5 & -0.1 & 0.7 & 0 \\  	
		& 2 & 0.98(0.18) & 0.19(0.02) & 0.39(0.02) & 0.84 & 0.39 & 0.5 & -0.1 & 0.7 & 0 \\  
		& 3 & 0.99(0.17) & 0.19(0.02) & 0.39(0.02) & 0.84 & 0.39 & 0.5 & -0.1 & 0.7 & 0 \\  
		\hline
		\multirow{3}{*}{3}  
		& 1 & 0.67(0.02) & 0.24(0.003) & 0.38(0.002) & 2.91 & 0.22 & 0 & -0.1 & 0 & 1 \\  
		& 2 & 0.67(0.02) & 0.24(0.003) & 0.38(0.002) & 2.91 & 0.22 & 0 & -0.1 & 0 & 1 \\  
		& 3 & 0.67(0.02) & 0.24(0.003) & 0.38(0.002) & 2.91 & 0.22 & 0 & -0.1 & 0 & 1 \\  
		\hline
		\multirow{3}{*}{4}  
		& 1 & 1.00(0.17) & 0.19(0.02) & 0.39(0.03) & 0.93 & 0.18 & 0.5 & -0.1 & 0.5 & 0.3 \\  
		& 2 & 1.00(0.17) & 0.19(0.02) & 0.39(0.03) & 0.93 & 0.18 & 0.5 & -0.1 & 0.5 & 0.3 \\  
		& 3 & 1.00(0.17) & 0.19(0.02) & 0.39(0.03) & 0.93 & 0.18 & 0.5 & -0.1 & 0.5 & 0.3 \\  
		\hline
		\multirow{3}{*}{5}  
		& 1 & 0.90(0.20) & 0.2(0.02) & 0.38(0.03) & 0.24 & 0.52 & 0.7 & -0.2 & 0.4 & 0.4 \\  
		& 2 & 0.90(0.20) & 0.2(0.02) & 0.38(0.03) & 0.24 & 0.52 & 0.7 & -0.2 & 0.4 & 0.4 \\  
		& 3 & -0.01(0.17) & 0.27(0.02) & 0.27(0.02) & 0.56 & 0.56 & 0.7 & -0.2 & 0 & 0 \\  
		\hline
		\multirow{3}{*}{6}  
		& 1 & 0(0.19) & 0.30(0.03) & 0.30(0.03) & 1.15 & 0.002 & 0.5 & 0 & 0 & 0 \\  
		& 2 & 0(0.19) & 0.30(0.03) & 0.30(0.03) & 1.15 & 0.002 & 0.5 & 0 & 0 & 0 \\  
		& 3 & 0.94(0.19) & 0.23(0.02) & 0.43(0.03) & 1.04 & -0.17 & 0.5 & 0 & 0.4 & 0.4 \\  
		\hline
		\multirow{3}{*}{7}  
		& 1 & 0.42(0.2) & 0.28(0.03) & 0.37(0.03) & 0.90  & 0.94  & 0.5 & -0.1 & 0.5 & -0.5 \\  
		& 2 & 0.75(0.18) & 0.22(0.02) & 0.37(0.03) & 0.90 & 0.43 & 0.5 & -0.1 & 0.5 & 0 \\  
		& 3 & 1.19(0.19) & 0.21(0.02) & 0.46(0.03) & 0.98 & 0.18 & 0.5 & -0.1 & 0.5 & 0.5 \\  
		\hline
		\multirow{3}{*}{8}  
		& 1 & 0.81(0.17) & 0.21(0.01) & 0.38(0.04) & 1.28 & 1.11 & 0.5 & -0.4 & -0.2 & 1.2 \\ 
		& 2 & 1.06(0.18) & 0.2(0.01) & 0.42(0.04) & 1.14 & 1.16 & 0.5 & -0.4 & 0 & 1.2 \\
		& 3 & 1.43(0.18) & 0.2(0.01) & 0.51(0.04) & 1.14 & 1.18 & 0.5 & -0.4 & 0.2 & 1.2 \\ 
		\hline
		
	\end{tabular}
	\caption{\label{tab:simulations_scenarios} Parameters for simulation studies. LOR: log odds ratio, RR$_c$: response rate in the control arm, RR$_t$: response rate in the treatment arm, as computed across 1,000 simulation replications (for Scenario 1, the average is across 5,000 simulation replications).}
\end{table}	

%Table 2
\begin{table} [h!]
	\begin{tabular}{cc|ccc|ccc}%ccc}
		
		Scenario   &  Subtrial  &
		\multicolumn{2}{c}{95\% HDI for LOR} & Reduction & 
		\multicolumn{2}{c}{Power} & Increase \\ 
		& & ABBA & BIN & in 95\% HDI & ABBA & BIN & in power 
		\\\hline
		\multirow{3}{*}{1}   
		& 1 & -0.84 - 0.83 & -1.17 - 1.11 & 26\% & 0.020 & 0.014 & $\star$ \\ 
		& 2 & -0.84 - 0.84 & -1.16 - 1.11 & 27\% & 0.018 & 0.013 & $\star$ \\ 
		& 3 & -0.85 - 0.82 & -1.16 - 1.11 & 26\% & 0.020 & 0.016 & $\star$ \\  
		\hline
		\multirow{3}{*}{2}   
		& 1 & 0.25 - 1.8 & 0.04 - 2.12 & 24\% & 0.75 & 0.52 & 43\% \\ 
		& 2 & 0.22 - 1.8 & 0.01 - 2.10 & 23\% & 0.71 & 0.53 & 35\% \\ 
		& 3 & 0.22 - 1.8 & 0.007 - 2.09 & 23\% & 0.70 & 0.53 & 33\% \\ 
		\hline
		\multirow{3}{*}{3}   
		& 1 & -0.09 - 1.5 & -0.30 - 1.74 & 21\% & 0.41 & 0.26 & 57\%  \\ 
		& 2 & -0.10 - 1.5 & -0.33 - 1.71 & 21\% & 0.41 & 0.25 & 65\%  \\ 
		& 3 & -0.10 - 1.5 & -0.31 - 1.75 & 21\% & 0.40 & 0.25 & 59\%  \\ 
		\hline
		\multirow{3}{*}{4}   
		& 1 & 0.27 - 1.85 & 0.05 - 2.12 & 24\% & 0.78 & 0.54 & 44\%  \\ 
		& 2 & 0.26 - 1.84 & 0.02 - 2.10 & 24\% & 0.75 & 0.55 & 37\%  \\ 
		& 3 & 0.25 - 1.83 & 0.02 - 2.09 & 24\% & 0.74 & 0.54 & 38\%  \\ 
		\hline
		\multirow{3}{*}{5}   
		& 1 & 0.05 - 1.6 & -0.24 - 1.84 & 25\% & 0.56 & 0.31 & 78\%  \\ 
		& 2 & 0.03 - 1.6 & -0.26 - 1.83 & 25\% & 0.54 & 0.31 & 73\%  \\ 
		& 3 & -0.59 to 1.1 & -0.77 - 1.34 & 22\% & 0.097 & 0.08 & 23\%  \\ 
		\hline
		\multirow{3}{*}{6}   
		& 1 & -0.62 to 0.88 & -0.84 to 1.15 & 25\% & 0.06 & 0.05 & 27\%  \\ 
		& 2 & -0.63 to 0.88 & -0.87 to 1.14 & 25\% & 0.06 & 0.04 & 44\%  \\ 
		& 3 & -0.06 to 1.5 & -0.33 to 1.68 & 22\% & 0.45 & 0.23 & 96\%  \\  
		\hline
		\multirow{3}{*}{7}   
		& 1 & -0.13 to 1.4 & -0.35 to 1.61 & 21\% & 0.38 & 0.24 & 57\%  \\  
		& 2 & -0.02 to 1.6 & -0.22 to 1.83 & 22\% & 0.51 & 0.35 & 46\%  \\  
		& 3 & 0.18 - 1.8 & 0.04 - 2.06 & 18\% & 0.68 & 0.54 & 27\%  \\ 
		\hline
		\multirow{3}{*}{8}   
		& 1 & 0.13 - 1.8 & 0.002 - 2.07 & 21\% & 0.64 & 0.52 & 23\%  \\  
		& 2 & 0.32 - 1.9 & 0.12 - 2.17 & 21\% & 0.79 & 0.61 & 30\%  \\  
		& 3 & 0.6 - 2.2 & 0.33 - 2.35 & 20\% & 0.95 & 0.76 & 25\%  \\ 
		\hline
	\end{tabular}
	\caption{\label{tab:results_basket}Mean 95\% HDI for the log odds ratios and one-sided power for subtrials in a basket trial. $^\star$In Scenario 1 (null scenario),  a type I error rate is controlled at 5\% for both the ABBA and the BIN methods. } 
\end{table}

% Table 3
\begin{table} [h!]
	\begin{tabular}{cc|ccc|ccc}
		Scenario  & Subtrial &
		\multicolumn{2}{c}{95\% HDI for LOR} & Reduction & 
		\multicolumn{2}{c}{Power} & Increase  \\ 
		& & ABBA & BIN & {in 95\% HDI} & ABBA & BIN & in power
		\\\hline
		\multirow{3}{*}{1} 
		& 1 & -1.06 to 1.04 & -1.66 to 1.47 & 33\% &0.033 & 0.024 & $\star$ \\
		& 2 & -1.04 to 1.05 & -1.34 to 1.47 &33\% & 0.034 & 0.024 & $\star$ \\  
		& 3 & -1.06 to 1.03 & -1.64 to 1.45 & 32\% &0.035 & 0.024 & $\star$ \\ 
		\hline
		\multirow{3}{*}{2}   
		& 1 & 0.1 - 2 & -0.22 - 2.5 & 28\% & 0.58 & 0.34 & 69\% \\  
		& 2 & 0.09 - 2 & -0.26 - 2.46 & 28\% & 0.57 & 0.35 & 63\% \\  
		& 3 & 0.07 - 2 & -0.28 to 2.42 & 28\% & 0.54 & 0.34 & 62\% \\  
		\hline
		\multirow{3}{*}{3}   
		& 1 & -0.19 to 1.7 & -0.51 - 2.05 & 26\% & 0.34 & 0.22 & 52\% \\
		& 2 & -0.21 to 1.7 & -0.58 to 2.01 & 26\% & 0.32 & 0.18 & 75\% \\  
		& 3 & -0.21 to 1.7 & -0.55 to 2.04 & 27\% & 0.33 & 0.2 & 63\% \\
		\hline
		\multirow{3}{*}{4}   
		& 1 & 0.13 - 2.1 & -0.2 to 2.51 & 28\% & 0.6 & 0.37 & 63\% \\ 
		& 2 & 0.11 - 2.1 & -0.26 to  2.46 & 28\% & 0.59 & 0.35& 68\% \\  
		& 3 & 0.1 - 2& -0.27 to 2.41 & 27\% & 0.57 & 0.35 & 61\% \\ 
		\hline
		\multirow{3}{*}{5}   
		& 1 & 0.05 - 2 & -0.31 to 2.37 & 29\% & 0.52 & 0.32 & 65\% \\ 
		& 2 & 0.02 - 1.9 & -0.36 to 2.32 & 28\% & 0.51 & 0.3 & 70\% \\ 
		& 3 & -1 to 0.97 & -1.37 to 1.21 & 23\% & 0.04 & 0.022 & 82\% \\ 
		\hline
		\multirow{3}{*}{6}  
		& 1 & -0.89 to 0.91 & -1.25 to 1.22 & 27\% & 0.043 & 0.032 & 34\% \\ 
		& 2 & -0.9 - 0.91 & -1.29 - 1.2 & 27\% & 0.046 & 0.032 & 44\% \\ 
		& 3 & 0.06 - 1.9 & -0.25 to 2.23 & 25\% & 0.54 & 0.35 & 53\% \\ 
		\hline
		\multirow{3}{*}{7}   
		& 1 & -0.45 to 1.3 & -0.76 to 1.74 & 26\% & 0.17 & 0.1 & 63\% \\ 
		& 2 & -0.14 to 1.7 & -0.5 to 2.13 & 28\% & 0.4 & 0.24 & 68\% \\ 
		& 3 & 0.33 - 2.2 & -0.006 to 2.58 & 26\% & 0.73 & 0.49 & 48\% \\ 
		\hline
		\multirow{3}{*}{8}   
		& 1 & -0.07 to 1.9 & -0.4 - 2.22 & 25\% & 0.44 & 0.25 & 74\% \\ 
		& 2 & 0.2 - 2.2 & -0.16 to 2.49 & 26\% & 0.65 & 0.41 & 60\% \\ 
		& 3 & 0.56 - 2.5 & 0.25 - 2.83 & 25\% & 0.87 & 0.67 & 30\% \\ 
		\hline
	\end{tabular}
	\caption{\label{tab:results_stratified}Mean 95\% HDI for the log odds ratios and one-sided power for the stratified analysis. $^\star$In Scenario 1 (null scenario),  a type I error rate is controlled at 5\% for both the ABBA and the BIN methods. 
	}
\end{table}

%Table 4

\begin{table}[h!]
	\begin{tabular}{c|c|c|c}
		Scenario   &  Subtrial   & Reduction & Increase \\ 
		& & in 95\% HDI & in power 
		\\ \hline
		\multirow{3}{*}{1}  
		& 1 & 20\% & $\star$ \\ 
		& 2 & 20\% & $\star$ \\ 
		& 3 & 20\% & $\star$ \\ 
		\hline
		\multirow{3}{*}{2}  
		& 1 & 18\% & 30\% \\  
		& 2 & 18\% & 26\% \\ 
		& 3 & 18\% & 29\% \\ 
		\hline
		\multirow{3}{*}{3}  
		& 1 & 15\% & 21\% \\ 
		& 2 & 15\% & 27\% \\ 
		& 3 & 15\% & 21\% \\ 
		\hline
		\multirow{3}{*}{4}  
		& 1 & 19\% & 31\% \\ 
		& 2 & 19\% & 28\% \\ 
		& 3 & 19\% & 31\% \\ 
		\hline
		\multirow{3}{*}{5}  
		& 1 & 18\% & 7\% \\ 
		& 2 & 18\% & 6\% \\ 
		& 3 & 17\% & 140\% \\ 
		\hline
		\multirow{3}{*}{6}  
		& 1 & 17\% & 44\% \\ 
		& 2 & 17\% & 22\% \\ 
		& 3 & 16\% & -17\% \\ 
		\hline
		\multirow{3}{*}{7}  
		& 1 & 13\% & 120\% \\ 
		& 2 & 15\% & 28\% \\ 
		& 3 & 14\% & -7\% \\ 
		\hline
		\multirow{3}{*}{8}  
		& 1 & 17\% & 47\% \\ 
		& 2 & 17\% & 22\% \\ 
		& 3 & 17\% & 9\% \\ 
		\hline
		
	\end{tabular}
	\caption{\label{tab:ab_abba} Reduction in 95\% HDI for the log odds ratios and increase in power for the ABBA method with information sharing vs the stratified ABBA methods. $^\star$In Scenario 1 (null scenario),  a type I error reduction of 40\%, 48\% and 40\% was achieved with information sharing compared to no sharing. }
\end{table}	

%Table 5
\begin{table} [h!]
	\begin{tabular}{c|cc|cc|cc|cc}
		& \multicolumn{4}{c|}{Trial  NCT01645280 (N = 151) } &  \multicolumn{4}{c}{Trial  NCT01077362 (N = 268) } \\ 
		\cline{2-9}
		& \multicolumn{2}{c|}{Control (N = 50)} 
		& \multicolumn{2}{c|}{Treatment (N = 101)} 
		& \multicolumn{2}{c|}{Control (N = 81)} 
		& \multicolumn{2}{c}{Treatment (N = 187)}\\
		& BinY & BinN & BinY & BinN  & BinY & BinN & BinY & BinN \\\hline
		\multirow{2}{*}{}   
		DAS28-CRP $<$ 2.6 & {\bf 5} & 0 & {\bf 16} & 0 & {\bf 28} & 1 & {\bf 70} & 3\\ 
		DAS28-CRP $\geq$ 2.6 & 42 & 3 & 79 & 6 & 45 & 7 & 96 & 18\\ \hline
		\multirow{2}{*}{}   
		ACR-N $>$ 0.2 & {\bf 25}  & 1 & {\bf 50} & 1  & {\bf 29} & 1 & {\bf 87} & 6\\ 
		ACR-N $\leq$ 0.2 & 22 & 2 & 45 & 5 & 44 & 7 & 79 & 15\\ \hline
	\end{tabular}
	\caption{\label{tab:realdata} Number of samples in the exemplar clinical trials, tabulated by the binary indicator and threshold for the DAS28-CRP measure. The numbers in bold represent the numbers of responders for the analysis with the binary model.}
\end{table}

%Table 6
\begin{table} [h!]
	\begin{tabular}{c|c|ccc|ccc}
		& & \multicolumn{3}{c|}{Stratified analysis } &  \multicolumn{3}{c}{Combined analysis }  \\
		\cline{3-8}	
		Outcome & Trial   
		& \multicolumn{2}{c}{95\% HDI} & $\Delta$ 95\% HDI  
		& \multicolumn{2}{c}{95\% HDI} & $\Delta$ 95\% HDI  \\
		& & ABBA & BIN & & ABBA & BIN &  \\\hline
		\multirow{2}{*}{DAS28}
		\multirow{2}{*}{}   
		& NCT01645280  & -0.13 to 1.02 & -0.46 to 1.7 & 46\% & 0.005 - 1.096 & -0.36 to 1.41 & 38\% \\ 
		& NCT01077362  & -0.20 to 0.51 & -0.39 to 0.66 & 31\% & -0.21 to 0.455 & -0.33 to 0.65 & 32\% \\ \hline
		\multirow{2}{*}{ACR-N}
		\multirow{2}{*}{}   
		& NCT01645280  & -0.44 to 0.63  & -0.72 to 0.63  & 21\% & -0.36 to 0.67 & -0.62 to 0.66 & 18\% \\ 
		& NCT01077362  & 0.15 - 1.00  & 0.03 - 1.05  & 16\% & 0.12 to 0.95 & -0.04 to 0.98 & 18\% \\ \hline
		
	\end{tabular}
	\caption{\label{tab:resultsreal} 95\% credible interval for the log odds ratios for the real data analysis. The $\Delta$ 95\% HDI represents the percentage decrease in the width of the 95\% HDI for the ABBA model relative to the BIN model. For NCT01645280 and NCT01077362, sharing of information in the ABBA model resulted in a 5\% and 6\% decrease in the width of the 95\% HDI for the DAS28 outcome, and a 4\% and 2\% decrease for the ACR-N outcome, respectively.}
\end{table}

%Figure1

%Figure1

\end{document}